\documentclass[aps,prd,10pt,twocolumn,preprintnumbers,superscriptaddress,nofootinbib]{revtex4-2}
\usepackage{amsmath,amssymb}
\usepackage{graphicx}
\usepackage{xcolor}
\usepackage[indent]{parskip}
\usepackage[stretch=15]{microtype}
\usepackage[normalem]{ulem}
\usepackage{hyperref}
\hypersetup{colorlinks,allcolors=[rgb]{0 0 0.75}}

\newcommand{\colorindicator}[1]{{\textcolor[HTML]{#1}{\scalebox{0.8}{$\blacksquare$}}}}

\newcommand{\BH}{{\mathrm{BH}}}
\newcommand{\dd}{{\mathrm{d}}}
\renewcommand{\vec}[1]{{\mathbf{#1}}}

\newcommand{\hl}{}

\makeatletter
\def\bibsection{%
   \par
   \begingroup
    \baselineskip26\p@
    \bib@device{\hsize}{72\p@}%
   \endgroup
   \nobreak\@nobreaktrue
   \addvspace{19\p@}%
  }%
\makeatother

\begin{document}
\preprint{\texttt{LAPTH-013/25}}
\title{Tidal effects on primordial black hole capture in neutron stars}
\author{Ian Holst}
\email{holst@uchicago.edu}
\affiliation{Department of Astronomy \& Astrophysics and Kavli Institute for Cosmological Physics, University of Chicago, Chicago, Illinois 60637, USA}
\author{Yoann G\'enolini}
\email{yoann.genolini@lapth.cnrs.fr}
\author{Pasquale Dario Serpico}
\email{serpico@lapth.cnrs.fr}
\affiliation{LAPTh, CNRS, USMB, F-74940 Annecy, France}
\date{\today}

\begin{abstract}
We revisit the problem of the capture of a primordial black hole (PBH) by a neutron star, accounting for the tidal perturbation from a nearby star or planet. For asteroid-mass PBHs, which could constitute all of the dark matter in the universe, a weakly bound post-capture orbit could be tidally disturbed to the point of preventing the PBH from settling in the neutron star and consuming it within a cosmologically short timescale. We show how this effect depends on environmental parameters and can weaken the proposed constraints based on observations of old neutron stars in high-density dark matter environments for PBH masses $\lesssim 10^{22}\,$g. We also provide approximate analytical formulae for the capture rates.
\end{abstract}

{
\parskip=0pt
\maketitle
}

\section{Introduction} \label{sec:intro}

Despite being among the first dark matter (DM) candidates to be proposed, primordial black holes (PBHs)~\cite{ZeldovichHypothesisCoresRetarded1966,HawkingGravitationallyCollapsedObjects1971,ChaplineCosmologicalEffectsPBHs1975} still remain viable in the present day. Even though they only experience gravitational interactions, they offer a surprisingly diverse set of detectable astrophysical signatures. In particular, PBHs could still account for 100\% of DM in the so-called asteroid-mass window within $[10^{-16},10^{-10}]\,M_\odot$. This range is constrained on the lower end by the condition that PBHs must survive to the present day without evaporating via Hawking radiation, and on the upper end by the non-observation of gravitational lensing events (see e.g. \cite{CarrConstraintsPBHs2020,Green:2020jor} for reviews). 

Potential signatures and constraints in the still unconstrained mass range involve interactions between PBHs and compact stars --- white dwarfs (WD) or neutron stars (NS) --- in DM-rich environments~\cite{CapelaConstraintsPBHDMNeutronStar2013,CapelaConstraintsPBHDMStarFormation2013,PaniTidalCapturePBH2014,GrahamDMTriggersSupernovae2015,CaiozzoRevisitingPBHCapture2024,Esser:2025pnt}. In the case of neutron stars, their compactness and density facilitate gravitational capture. Once a PBH is trapped, it gradually grows and ultimately consumes the star, ``transmuting'' it into a black hole. This dramatic process is expected to produce detectable signatures across various messengers and wavelengths. There is no consensus on a fiducial model for this phenomenon, but it likely manifests as a radio burst~\cite{FullerDMinducedCollapseNeutron2015,Fuller:2017uyd,AbramowiczCollisionsNeutronStars2018}, kilonova-like photon signal~\cite{Fuller:2017uyd,BramanteSearchingDMNeutron2018}, positrons~\cite{Fuller:2017uyd,TakhistovPositronsPBHMicroquasars2019}, and gravitational waves~\cite{TakhistovTransmutedGravityWave2018,AbramowiczCollisionsNeutronStars2018,KuritaGWsDMCollapse2016,BaiottiGravitationalRadiationCollapse2007}. In this context, numerical simulations offer a promising path forward, though current efforts are largely limited to general relativistic hydrodynamical (GRHD) simulations, as in \cite{EastFateNeutronStar2019,Baumgarte:2024ouj}. These approaches do not yet capture the full complexity of NS physics, notably the effects of intense magnetic fields. Nevertheless, they have yielded valuable insights, including the stability of the NS when traversed by a sufficiently low-mass PBH, as well as the generation of shock waves in the PBH’s wake.

Although the signatures remain uncertain, the \textit{rate} of the PBH capture leading to transmutation can be more reliably computed \cite{CapelaConstraintsPBHDMNeutronStar2013,GenoliniRevisitingPBHCapture2020}, provided that the traditional assumption that the PBH--NS system is isolated holds. However, even if the first passage of a PBH near or through an NS results into a capture, the PBH is typically weakly bound and on a highly eccentric orbit. As highlighted in~\cite{MonteroCamachoRevisitingConstraintsAsteroidmass2019}, under these conditions and particularly for low-mass PBHs, tidal perturbations could substantially alter the orbit, to the point of preventing the PBH from settling in the NS and consuming it within a timescale that is short with respect to the lifetime of the universe, or of the system of interest. In this paper, we attempt to assess more precisely the critical PBH masses below which stellar and planetary perturbations disrupt the standard calculations, making them unreliable and suppressing the rate of captures that would lead to most of the interesting signatures. We shall also illustrate how these results depend on the properties of the DM and stellar environments, as well as on the still poorly known statistics of planetary systems around NS.

This article is structured as follows: In Sec.~\ref{sec:capture}, we review the basics of PBH capture in NS, which also gives us the opportunity to introduce key notations. Sec.~\ref{sec:tidal_stars} outlines our approach to accounting for the tidal interaction of the nearest perturber external to the PBH--NS binary. Sec.~\ref{sec:effects} is devoted to the effects on the relevant rates, where the parametric dependence on the environmental properties (such as star number density or velocity dispersion) is presented. In Sec.~\ref{sec:tidal_planets} we extend the formalism --- to the best of our knowledge, for the first time --- to the case of planetary perturbations. \hl{Some considerations on its applicability and limitation to the case in which the NS has a \textit{stellar} companion are reported in Sec~\ref{sec:tidal_stars_bin}.} A discussion of the results obtained and conclusions are reported in Sec.~\ref{sec:discussion}. Some technicalities related to the motion and energy losses of the PBH within the NS are detailed in Appendices \ref{appendix:interior_orbits} and \ref{appendix:e_loss}. In addition, in Appendix~\ref{appendix:analytic} we derive approximate analytical formulae for the capture and merger rates, which are useful to gauge the scalings with different parameters.

\section{PBH Capture in Neutron Stars} \label{sec:capture}

We first review the scenario of a light PBH--NS interaction without additional bodies. Consider a PBH of mass $m$ which approaches a NS of mass $M$ on an unbound, hyperbolic orbit. We assume $m \ll M$ so that the NS is considered at rest in the barycentric reference frame. \hl{Furthermore, for the phases well before the PBH settles in the NS, the non-relativistic approximation holds}~\cite{Baumgarte:2024iby,Baumgarte:2024ouj}. The initial hyperbolic orbit can be parametrized with two variables: an initial impact parameter, $b_0$ (geometrically, this is equivalent to the semi-minor axis of the hyperbola), and the initial speed at infinite separation, $v_0$ (see Fig.~\ref{fig:schematic}).

Given these initial parameters, the PBH has an initial total energy and initial angular momentum about the NS
\begin{equation}
    E_0 = \frac{1}{2} m v_0^2 \quad \text{and} \quad \vec{L}_0 = m v_0 b_0 \hat{\vec{z}},
\end{equation}
respectively, $\hat{\vec{z}}$ being the unit vector orthogonal to the orbital plane. The initial semi-major axis of the hyperbolic orbit is defined
\begin{equation}
    a_0 = \frac{G M}{v_0^2} = \frac{r_s}{2} \left(\frac{c}{v_0}\right)^2 \;,
\end{equation}
where $r_s$ is the Schwarzschild radius of the NS. The orbit has an eccentricity $e_0 = \sqrt{1 + (b_0 / a_0)^2}>1$.

In a polar coordinate system centered on the NS with azimuthal angle $\phi$, the initial hyperbolic orbit of the PBH is described by
\begin{equation}
    r_\mathrm{hyp}(\phi) = \frac{b_0^2}{a_0 + \sqrt{a_0^2 + b_0^2} \cos{\phi}},
\end{equation}
where $\phi$ is restricted to be $|\phi| < \cos^{-1}(-1/e_0)$. Since the motion is in a plane, we parametrize the orbital position vector in Cartesian coordinates as
\begin{equation}
    \vec{r} = r_\mathrm{hyp}(\phi) \left(\begin{smallmatrix}
        \cos\phi\\
        \sin\phi\\
        0
    \end{smallmatrix}\right)\,.
\end{equation}
The PBH enters the interior of the NS (which has radius $R$) during its first closest approach if $b_0$ is below the critical impact parameter $b_c$:
\begin{equation} \label{eq:bc}
   b_0 < b_c \equiv R \sqrt{1 + \frac{2 a_0}{R}}\,.
\end{equation}
When this condition is satisfied, dissipative interactions are sizable\footnote{Actually, sizable dissipation effects can also affect PBHs passing outside the NS at their closest approach, via gravitational wave emission. However, this is only relevant for masses above the range affected by tidal perturbations, as we will show below.} and the gravitational potential felt by the PBH is significantly altered, both effects resulting in major modifications of the orbital path.

If the NS has a constant density profile (an approximation valid within $\sim 10\%$ for the majority of the NS volume, see e.g. Ref.~\cite{GenoliniRevisitingPBHCapture2020}) the PBH trajectory within the NS can be derived analytically, as discussed in Appendix~\ref{appendix:interior_orbits} in more detail. The main consequences are that the maximum speed of the PBH at periapsis is lower than it would be in vacuum, and an overall orbital precession is induced by the deviation from a Keplerian trajectory. In particular, this second effect complicates the introduction of a tidal perturbation because it causes the orientation of the orbit to change (eventually, in a chaotic way) over time.

To be captured into a bound elliptical orbit during its initial passage near the NS, the PBH must lose an energy $|\Delta E| > E_0$ through dissipative interactions. The energy losses we account for are reviewed in Appendix \ref{appendix:e_loss}; there we show that while energy losses are largely dominated by accretion and dynamical friction for $b_0<b_c$, gravitational waves are the only relevant dissipative process for $b_0>b_c$. As shown in Fig.~3 of Ref.~\cite{GenoliniRevisitingPBHCapture2020}, the proportion of captures that occur in either case varies significantly with the mass and the initial dispersion speed of the PBHs. Assuming a typical DM dispersion speed $\sigma$, gravitational wave capture dominates for PBH masses above $\sim 10^{28}\,\mathrm{g} \times (\sigma \, / \, 10^{-3} c)^2$.

If the PBH becomes bound to the NS as a consequence of these energy losses, then upon exiting the NS, the semi-major and semi-minor axes of the first elliptical orbit are
\begin{align}
    a_1 &= -\frac{G M m}{2 E_1}\quad \text{with} \quad E_1=E_0 + \Delta E <0 \,,\\
    b_1 &= \frac{|\vec{L}_1|}{\sqrt{-2m E_1}}\quad \text{with} \quad \vec{L}_1=\vec{L}_0 + \Delta \vec{L}\,.
\end{align}
The eccentricity is then $e_1 = \sqrt{1 - (b_1 / a_1)^2} < 1$, and the bound elliptical orbit is defined in polar coordinates by
\begin{equation}
    r_\mathrm{ell}(\phi) = \frac{b_1^2}{a_1 + \sqrt{a_1^2 - b_1^2} \cos{\phi}},
\end{equation}
where $0 \ge \phi > 2\pi$. Note that, in practice, $|\Delta \vec{L}| \ll |\vec{L}_0|$, as can be seen in Figure~\ref{fig:energy_losses} of Appendix~\ref{appendix:e_loss}, so one can usually approximate $\vec{L}_1 \approx \vec{L}_0$.

\section{Tidal Perturbations from Stars} \label{sec:tidal_stars}

\begin{figure}
    \centering
    \includegraphics[width=\linewidth]{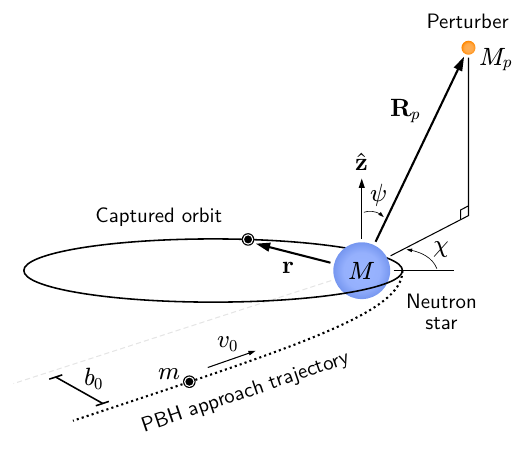}
    \caption{Diagram of the PBH--NS--perturber system illustrating the geometry and relevant free parameters.}
    \label{fig:schematic}
\end{figure}

We now add a third body, typically a neighboring star with mass $M_p$ at distance $R_p$ from the NS. This perturber is oriented with polar angle $\psi$ and azimuthal angle $\chi$, defined relative to the periapsis and plane of the first captured PBH orbit, as schematically shown in Figure~\ref{fig:schematic}. The Cartesian position vector of the perturber is then
\begin{equation}
    \vec{ R}_p = {R}_p \left(\begin{smallmatrix}
        \sin\psi \cos\chi\\
        \sin\psi \sin\chi\\
        \cos\psi
    \end{smallmatrix}\right)\,.
\end{equation}

This 3-body problem involving the PBH, NS, and perturber is not solvable in full generality, so we will resort to some approximations appropriate for the situation of interest. Of primary interest is the case where the PBH dynamical timescale after capture is much shorter than the timescale over which the perturber moves significantly. If the position of the perturber is assumed to be constant during the period of the captured orbit, then given that the perturber likely has a similar mass to the NS, this is approximately equivalent to requiring that the distance of the perturber is much larger than the typical NS--PBH orbital distance ($R_p \gg a_1$).
\hl{This stationary perturber approximation is valid for PBH masses $m \gg 2 \times 10^{19}\,\mathrm{g} \times (\mathrm{pc} / R_p)$, which can be derived from Eq.~\eqref{eq:deltaE_approx} and Eq.~\eqref{eq:a1_approx} in Appendix~\ref{appendix:analytic}.}

The tidal force due to the perturber is related to the relative gravitational acceleration of the PBH and NS:
\begin{equation}
\begin{split}
    \vec{F}_t &= m (\vec{a}_\BH - \vec{a}_\mathrm{NS})\\
    &= G M_p m \left( \frac{\vec{R}_p - \vec{r}}{|\vec{R}_p - \vec{r}|^3} - \frac{\vec{R}_p}{|\vec{R}_p|^3} \right) \\
    &\approx \frac{G M_p m}{R_p^3} \left( \frac{3 (\vec{r} \cdot \vec{R}_p)}{R_p^2} \vec{R}_p - \vec{r} \right).\label{TidalForceApprox}
\end{split}
\end{equation}
In the last line (and henceforth) we assume $r \ll R_p$ which is generally required to treat the perturber as stationary, and makes further calculations analytically tractable. The tidal force induces a torque on the PBH
\begin{equation}
    \boldsymbol{\tau} = \vec{r} \times \vec{F}_t =\frac{3 G M_p m}{R_p^5} (\vec{r} \cdot \vec{R}_p) (\vec{r} \times \vec{R}_p),\label{torquep}
\end{equation}
which causes a change in the orbital angular momentum. During one full elliptical orbit with semi-major axis $a$ and eccentricity $e$, this change in $\vec{L}$ amounts to
\begin{align} \label{eq:delta_L}
    \delta \vec{L} &= \int_0^{T_1} \dd t \, \boldsymbol{\tau} = \int_0^{2\pi} \dd\phi \, \left(\frac{m r^2}{L}\right) \boldsymbol{\tau} \\
    &= \frac{3 \pi G M_p m}{\sqrt{G M / a}} \left(\frac{a}{R_p}\right)^{\! 3} \begin{pmatrix}
        (1-e^2) \sin\psi \cos\psi \sin\chi\\
        -(1 + 4e^2)\sin\psi \cos\psi \cos\chi\\
        5e^2 \sin^2\psi \sin\chi \cos\chi
    \end{pmatrix}\,,\nonumber
\end{align}
where we have expressed the angular momentum of the elliptical orbit as $L = m\sqrt{G M a (1-e^2)}$, in terms of the semi-major axis and eccentricity. Note that we use $\delta$ symbols for changes of $E$ or $L$ due to tidal perturbation, while the notation $\Delta$ is used for changes due to PBH passage through or near the NS.

For this step, like when computing the energy losses, we are adopting a perturbative approach, where the change in angular momentum $\delta \vec{L}$ is integrated over the unperturbed trajectory. Consistently, $\delta E$ must be $0$. This is by construction, since we assume that the perturber is stationary and thus the work done by a conservative gravitational field along a closed trajectory is 0.

We compute the new orbital parameters, $a_1'$ and $b_1'$, as a result of the tidal perturbation:
\begin{align}
    a_{1}' &= a_{1} \quad \text{since} \quad E_1'=E_1\,,\\
    b_{1}' &= \frac{|\vec{L}'_1|}{\sqrt{-2m E_1'}} \quad \text{with} \quad \vec{L}_1'= \vec{L}_1 + \delta\vec{L} \,.
\end{align}
The perturbed orbital parameters $a_1'$ and $b_1'$ are then used to compute the trajectory of the PBH returning to the subsequent closest approach near or inside the NS (including a possible portion inside the NS). The new energy loss $\Delta E_2$ is computed again using the methods described in Appendix~\ref{appendix:e_loss}. After this second encounter, the PBH exits from the vicinity or interior of the NS with altered orbital parameters:
\begin{align}
    a_{2} &= -\frac{G M m}{2 E_2}, \quad \text{with} \quad E_2=E_1' + \Delta E_2\,, \\
    b_{2} &= \frac{|\vec{L}_2 |}{\sqrt{-2m E_2}} \quad \text{with} \quad \vec{L}_2=\vec{L}_1' + \Delta \vec{L}_2\,.
\end{align}

In principle the iterative scheme with $\Delta E_i$ and $\delta \vec{L}_i$,
\begin{equation*}
    (a_i,b_i)\to (a_i',b_i')\to(a_{i+1},b_{i+1})\to (a_{i+1}',b_{i+1}') \to \dots
\end{equation*}
could be extended to further orbits. However, beyond the second pass through the NS, we ignore further effects of the perturber, and only consider the effect of subsequent energy loss on the PBH orbit. This approach is justified since:
\begin{enumerate}
    \renewcommand{\theenumi}{\Roman{enumi}}
    \item The effects of the perturber over multiple orbits do not add up coherently; because of the precession due to passage inside the NS, the orientation of the perturber relative to subsequent orbits changes, in practice in a chaotic manner.
    \item Unless the PBH orbit is disrupted early on by a significant effect of the perturber, as a consequence of repeated $E-$losses, the orbits tend to circularize. Hence, the PBH tends to suffer more moderate tidal perturbations, since the largest contribution to the torque scales quadratically with the apoapsis (see Eq.~\eqref{torquep}).
    % \item Although for specific orbits this approximation may occasionally lead to an error in the fate of the PBH, when averaging over many configurations, we expect that the false positives and false negatives should compensate, only affecting the average at a sub-leading level. Since we are seeking only a statistical answer to the problem, making a judgment when truncating the effect of the perturber after its first-orbit effect should be acceptable.
\end{enumerate}
\hl{We note that, in rare cases, our approximations are expected to fail: for instance, there might be circumstances where the accumulation of small tidal perturbations over many orbits could counteract the effect of energy losses and move the periapsis outwards. This is a ``false negative case'', where our estimate would suggest no disruption, but in fact there is one. Or there may be ``false positive cases'', where genuine three-body dynamics is actually responsible for a capture followed by a merger, while the simplified perturbed two-body treatment yields none. While an estimate of the size of these (small) effects goes beyond the goals of this article, we note that their role could be mitigated by the fact that the actual error on the \textit{statistical} quantities of interest here is due to the net effect of false positives and negatives, and not by the sum of their absolute errors.}
%\ih{we expect our approximation to produce both false positives and false negatives for various configurations in parameter space} 
%\hl{For example}, there might be contrived circumstances where the accumulation of small tidal perturbations over many orbits could counteract the effect of energy losses and move the periapsis outwards. Since we ignore this unlikely possibility, our results are, if anything, conservative with respect to the range of masses where tidal effects are important.\ih{this was an example of a false positive, should we add an example of a false negative too? I can't actually think of any}

Our criterion for whether a PBH capture leads to a successful transmutation of the NS depends on whether the entire process, from initial encounter to transmutation, happens on a timescale smaller than the typical age of the system of interest, $t_\mathrm{age}$. In the case of the Milky Way, $t_\mathrm{age}\simeq 10^{10}$ yr. Actually, the majority of the time is taken up by the orbital inspiral, because once the PBH orbit is mostly or entirely inside the NS, the final settling and accretion stages leading up to transmutation are relatively fast for the relevant PBH mass range. Using a general relativistic (GR) framework, accretion in a NS with a stiff EoS has been recently reviewed in \cite{Richards:2021zbr,Richards:2021upu}. The engulfing time of the NS was computed and found to be $0.6\,\mathrm{yr} \times (m \, / \, 10^{22}\,\mathrm{g})^{-1}$ \cite{Schnauck:2021hlm, Baumgarte:2021thx}, which is very close to the rough estimate given in previous studies, considering the Bondi accretion regime with a typical sound speed in neutron dense matter \cite{CapelaConstraintsPBHDMNeutronStar2013,KouvarisGrowthBHsInterior2014,GenoliniRevisitingPBHCapture2020}.

The inspiral stage approximately concludes once the semi-major axis of the PBH orbit reaches the NS radius ($a \sim R$), or equivalently when its orbital energy is
\begin{equation}
    E_f = -\frac{GMm}{2R}\,.
\end{equation}
For an estimate of the relevant timescale, one can assume that energy losses are constant at each passage. To take into account any significant tidal effects during the first bound orbit, we use $\Delta E'$, from the second passage through the NS, as a proxy for this constant energy loss. Reaching the inspiral stage then requires a number of orbits given by
\begin{equation}
N_\mathrm{orb} \approx \frac{E_f - E_1}{\Delta E'}\,.
\end{equation}
Summing all of the orbital periods results in a total inspiral time
\begin{align}
T_\mathrm{ins} &= \sum_{j=1}^{N_\mathrm{orb}} T_j = \sum_{j=1}^{N_\mathrm{orb}} 2 \pi \sqrt{\frac{a_j^3}{G M}} \\
&= \sum_{j=1}^{N_\mathrm{orb}} 2 \pi G M \left[ -\frac{2}{m} \Big( E_1 + (j-1) \Delta E' \Big) \right]^{-3/2} \nonumber\\
&= \frac{2 \pi G M m^{3/2}}{(-2 \Delta E')^{3/2}} \left[\zeta_H\!\left(\frac{3}{2},\frac{E_1}{\Delta E'}\right) - \zeta_H\!\left(\frac{3}{2},\frac{E_f}{\Delta E'}\right)\right], \nonumber
\end{align}
where $\zeta_H$ is the Hurwitz zeta function. Because energy losses shorten the orbital period, the sum is dominated by the first few orbits, and the second term which depends on $E_f$ does not significantly affect the result.

The perturber may affect the outcome in two ways. In some cases, it can impart enough angular momentum to push the periapsis outside the NS, lengthening the inspiral time enough to fail our criterion. However, there are approximately just as many angular orientations where it has the opposite effect. More frequently, the perturber acts by entirely disrupting the PBH--NS bound system, in which case the criterion is automatically not met. We consider the system to be disrupted if $|\delta\vec{L}| > |\vec{L}_1|$. Technically, this breaks the perturbative assumption and a full three-body analysis is needed at that point, but it is reasonable to believe that non-perturbative three-body dynamics typically result in a significant delay in merger or disruption of the PBH orbit, which is statistically the most frequent outcome. Thus, if $|\delta\vec{L}| > |\vec{L}_1|$, then we assume $T_\mathrm{ins} \to \infty$. A slightly less stringent threshold for disruption would be the point where $|\delta\vec{L}|$ is large enough that $b_1' > a_1'$, but we find that this does not make a significant difference in our results.\footnote{When $|\delta\vec{L}| > |\vec{L}_1|$ is allowed, at least up until $b_1' = a_1'$, then the PBH mass below which the merger rate is perturber-suppressed decreases by a factor of $\sim 1.5$.}

\section{Effect on NS Transmutation Rates in Different Environments}\label{sec:effects}

We now move beyond the single PBH case treated above to consider a \textit{population} of PBHs comprising DM. To estimate the rate of NS transmutation, we must consider the phase space of initial PBH orbital parameters and the distribution of perturber properties. We assume a benchmark Maxwellian initial velocity distribution for the PBHs with 1D velocity dispersion $\sigma$:
\begin{equation}
    p(v_0) = \sqrt{\frac{2}{\pi}} \frac{v_0^2}{\sigma^3} \exp\left(-\frac{v_0^2}{2 \sigma^2} \right)\,.
\end{equation}
Note that the old NS of interest here are expected to have a low velocity dispersion~(see e.g.~\cite{1992ApJ...384..105F}), so for our estimates we neglect the subleading relative motion of the NS with respect to the halo.

In the rest of this section, we consider the perturber to be a nearby star. We take the distribution of perturber masses $M_p$ to be a Salpeter initial stellar mass function~\cite{Salpeter:1955it} with a power law slope $-\gamma$ running between $M_\mathrm{min}$ and $M_\mathrm{max}$
\begin{equation} \label{eq:pM_p}
    p(M_p) = \frac{(\gamma - 1)}{M_\mathrm{min}^{1-\gamma} - M_\mathrm{max}^{1-\gamma}} \, M_p^{-\gamma}\,.
\end{equation}
We take the distribution of distances $R_p$ to be the nearest neighbor distance distribution modeled as a Poisson process \cite{BallesterosMergerRatePBHs2018}, giving
\begin{equation} \label{eq:pR_p}
    p(R_p) = 4\pi n_p R_p^2 \, \exp{\left( -\frac{4}{3}\pi n_p R_p^3 \right)},
\end{equation}
where $n_p$ is the average number density of stars nearby. We include only the effect of the nearest neighbor because,
%in most cases of interest, the next-to-nearest neighbor and beyond are likely to have a much smaller perturbative effect. %This is because although the dependence of Eq.~\eqref{eq:delta_L} on $1/R_p^3$ is compensated by the growing volume engulfing more perturbers, the perturbations due to the growing number of sources largely cancel out due to the different, uncorrelated directions of the forces 
\hl{in the limit of sizable tidal perturbations responsible for the disruption, this is the dominating effect}~\cite{RevModPhys.15.1,2009EPJB...70..413C,2018MNRAS.474.1482P}. Finally, we take the angles specifying the perturber orientation, $\psi$ and $\chi$, to be drawn from an isotropic distribution. This is reasonable since the distance to the nearest star is typically much smaller than the extent of the stellar environment (e.g. the thickness of the galactic disk) in most astrophysical environments of interest.

Note the degeneracy of $M_p$ and $R_p$ in the expression for the tidal perturbation of the orbital angular momentum, $\delta \vec{L}$, in Eq.~\eqref{eq:delta_L}. Taking advantage of this to reduce the dimensionality of the phase space, we define the quantity
\begin{equation} \label{Ydef}
    Y \equiv \frac{M_p}{R_p^3}.
\end{equation}
The corresponding probability density of $Y$ is found by appropriate transformation of Eqs.~(\ref{eq:pM_p}) and (\ref{eq:pR_p}):
\begin{equation}
        p(Y) = \frac{(\gamma - 1)}{Y} \, \frac{\mathsf{\Gamma} \big(2-\gamma, \tfrac{Y_\mathrm{min}\mathstrut}{Y}\big)-\mathsf{\Gamma} \big(2-\gamma, \tfrac{Y_\mathrm{max}\mathstrut}{Y}\big)}{\big(\frac{Y_\mathrm{min}\mathstrut}{Y}\big)^{1-\gamma} - \big(\frac{Y_\mathrm{max}\mathstrut}{Y}\big)^{1-\gamma}}
\end{equation}
where $\mathsf{\Gamma}(n,x)$ is the upper incomplete gamma function, $Y_\mathrm{min} = \frac{4}{3}\pi n_p M_\mathrm{min}$, and $Y_\mathrm{max} = \frac{4}{3}\pi n_p M_\mathrm{max}$.

Integrating over phase space, the average rate of PBH capture leading to merger by present day is
\begin{align} \label{eq:rate}
    \Gamma_\mathrm{mrg} &= \frac{\rho}{m} \Bigg( \int\limits_{0}^{\infty} \dd v_0 \, p(v_0)  v_0\int\limits_{0}^{b_\mathrm{max}} \dd b_0 \, 2 \pi b_0 \int\limits_0^\infty \dd Y \, p(Y) \nonumber \\
    & ~~~~~~~~ \frac{1}{4\pi} \int\limits_0^{2\pi} \dd\chi \int\limits_0^\pi \dd\psi \sin\psi ~
    \Theta(t_\mathrm{age} - T_\mathrm{ins}) \Bigg),
\end{align}
where $\rho$ is the energy density of PBH dark matter and $\Theta$ is the Heaviside step function, which enforces the criterion of the inspiral time being less than the age of the system. We set the maximum impact parameter to the average distance between stars, $b_\mathrm{max} \sim {n_p}^{-1/3}$, although the capture probability typically becomes negligible already for much smaller values of $b_0$; hence the exact choice of $b_\mathrm{max}$ does not have any practical impact on our results. We use a Monte Carlo integrator to perform this integral \cite{Lepage:2020tgj}.

To illustrate how the results depend on the parameters, we target two environments: the Milky Way's galactic disk, and a typical dwarf galaxy. One significant difference between these two representative cases is their DM velocity dispersion. For the Milky Way we fix $\sigma$ to the fiducial value $5\times 10^{-4} c$. This value is obtained for an isothermal DM halo with a flat rotation curve with $230 \, \mathrm{km/s}$ circular velocity \cite{BinneyGalacticDynamics1987}. For the `typical' dwarf galaxy we take $\sigma = 3 \times 10^{-5}c$ \cite{Walker:2007ju}. For visualization purposes, and to highlight the effect of differing velocity dispersions, we keep the other parameters the same for both cases. We take the density of PBH dark matter to be a uniform $\rho = 1\,\mathrm{GeV}/\mathrm{cm}^3$. The number density of perturbers $n_p$ is the typical stellar number density. In the solar neighborhood, $n_p = 0.14\, \mathrm{pc}^{-3}$ \cite{Gregersen2009}, but it mildly grows towards the inner Galaxy, jumping abruptly by up to 7 orders of magnitude in the innermost parsec. To illustrate the dependence on this parameter, we show results for both $n_p = 0.1\,\mathrm{pc}^{-3}$ and $n_p = 1\,\mathrm{pc}^{-3}$. In the Salpeter initial mass function, we use $M_\mathrm{min} = 0.5\,M_\odot$, $M_\mathrm{max} = 10\,M_\odot$, and $\gamma = 2.3$ \cite{Kroupa:2002ky}. Since the oldest NS have an age comparable to the Milky Way, and NS formation probably peaked in the past, consistently with past literature we take the age of the environment, and thus the maximum observable inspiral time, to be $t_\mathrm{age} = 10~\mathrm{Gyr}$. Based on \cite{PotekhinAnalyticalRepresentationsUnified2013}, we take the NS mass $M = 1.52 \, M_\odot$ and NS radius $R = 11.6\,\mathrm{km}$.

\begin{figure*}
    \centering
    \includegraphics[width=0.495\linewidth]{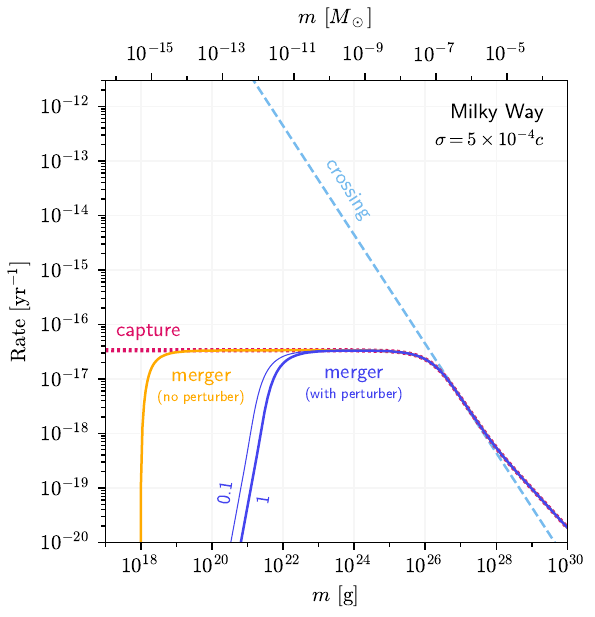}
    \includegraphics[width=0.495\linewidth]{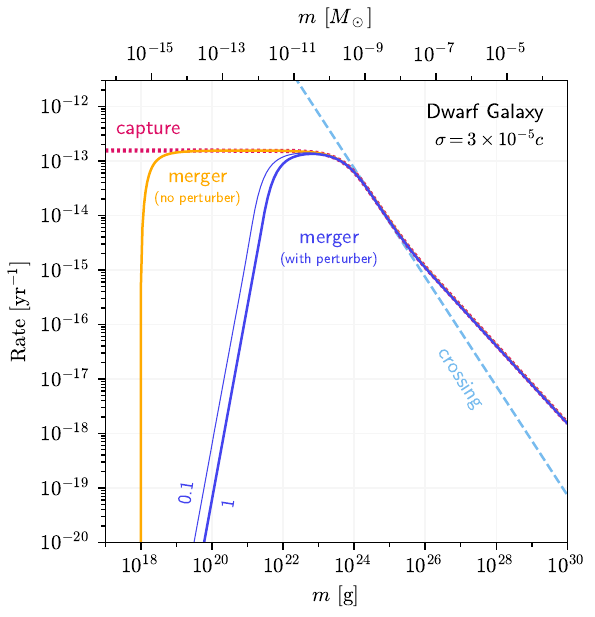}
    \caption{Rates for PBHs crossing through a NS (dashed light blue \colorindicator{77BBEE}), becoming captured (dotted red \colorindicator{DD1166}), merging by present day (yellow \colorindicator{FFAA00}), and merging by present day with the addition of a tidal perturber (blue \colorindicator{4444EE}). The two blue lines are calculated with different values of perturber number density $n_p$, indicated in units of pc$^{-3}$. The left panel is calculated using a velocity dispersion $\sigma$ appropriate for the Milky Way, and the right for a dispersion typical of dwarf galaxies.}
    \label{fig:rates}
\end{figure*}

In Figure~\ref{fig:rates} we show lines representing the rates of four different processes as a function of the PBH mass. Each line corresponds to the substitution of different criteria in the $\Theta$ function in Eq.~\eqref{eq:rate}. The \textit{merger (with perturber)} rate is precisely as defined in Eq.~\eqref{eq:rate}, while \textit{merger (no perturber)} is similar but neglecting the effect of the perturber and hence the associated integrals over $Y$, $\chi$, and $\psi$ trivially replaced by unity. \textit{Capture} refers to all encounters in which PBHs lose enough energy to become bound to the NS, satisfying the criterion $|\Delta E| > E_0$, even if they do not eventually lead to merger. \textit{Crossing} refers solely to cases in which the PBH passes inside the NS, satisfying $b_0 < b_c$. Neither capture nor crossing rates include the effect of a perturber.

Although the results of Figure~\ref{fig:rates} are obtained numerically, we confirm the perturbed merger rate curves (in blue) with approximate analytic expressions valid in four regimes of mass. Since these can be useful to rescale our results to different choices of parameters, the details of their derivation and the assumptions made are reported in Appendix~\ref{appendix:analytic}. In the following, we shall explain these four regimes, starting from high masses and going towards low masses.

At the highest masses, capture is dominated by gravitational waves. This is because energy losses grow with PBH mass, so once gravitational wave losses are large enough to lead to capture, more of the $b_0$--$v_0$ phase space is available for capture than there would be for contact-only interactions. In fact, we see that the capture and merger rates exceed the crossing rate in this regime, indicating that non-contact forces dominate. The gravitational wave capture rate is approximately 
\begin{equation}
    \Gamma_\mathrm{gw} \approx 14 \, \frac{G^2 M^{12/7} \rho}{c^{10/7} \sigma^{11/7} m^{5/7}}\,.
\end{equation}

At slightly lower masses, where gravitational wave emission is the subdominant mode of energy losses, the rates all match the benchmark ``crossing'' rate, meaning that all NS--PBH encounters will lead to capture and later, merger. The crossing rate is determined solely by orbital geometry and the NS radius $R$, and it can be written
\begin{equation}
    \Gamma_\mathrm{cross} \approx 5.0 \, \frac{G M R \, \rho}{\sigma m} \,.
\end{equation}

At even lower masses, passing through the NS does not always dissipate enough energy to capture the PBH, and the capture rate dips below the crossing rate, reaching a plateau at the value
\begin{align}
    \Gamma_\mathrm{plat} &\approx 1.5 \times 10^2 \, \frac{G^2 M \rho}{\sigma ^3}
    \\
    &\approx 3.4 \times 10^{-17} \, \mathrm{yr}^{-1}  \left(\frac{\rho}{\mathrm{GeV/cm^3}}\right) \! \left(\frac{5 \times 10^{-4}c}{\sigma}\right)^{\!3} \! . \nonumber
\end{align}

As energy losses decrease at lower masses, the captured orbits become larger and are more susceptible to disruption by a tidal perturber. This significantly suppresses the merger rate below a PBH mass of
\begin{equation} \label{eq:m_pert}
\begin{split}
    m_\mathrm{pert} &\approx 0.080 \, M^{5/7} R^{6/7} M_\mathrm{min}^{2/7} n_p^{2/7}
    \\
    &\approx 3.9 \times 10^{21}\,\mathrm{g} \, \left( \frac{M_\mathrm{min}}{0.5 \,M_\odot} \right)^{2/7}\left( \frac{n_p}{\mathrm{pc}^{-3}} \right)^{2/7}.
\end{split}
\end{equation}
Below this mass, the gradual slope of the merger rate results from the integrals over $\chi$ and $\psi$, in which a growing fraction of the perturber's angular phase space allows it to disrupt the PBH orbit. The merger rate in this regime (derived in Appendix~\ref{appendix:analytic}) is
\begin{equation} \label{eq:Gamma_pert}
    \Gamma_\mathrm{pert} \approx 2.8 \times 10^5 \, \frac{G^2 m^{7/2} \rho}{M^{3/2} R^3 \sigma^3 M_\mathrm{min} n_p}\,.
\end{equation}
Note that both $m_\mathrm{pert}$ and $\Gamma_\mathrm{pert}$ depend on the perturber number density $n_p$, since this controls the typical distance to the nearest perturber and thus the strength of the effect.

It is important to note that although our calculations show the transmutation rate plummeting for lower PBH masses, at some point there will in fact be small but finite contributions from more elaborate three-body interactions which still might eventually lead to stable capture. However, even in the absence of a perturber, there is still a mass cutoff below which the inspiral time is longer than the typical NS age, making it implausible for this process to cause any NS transmutations by present day (yellow line in Fig.~\ref{fig:rates}). This cutoff mass can be parametrized as
\begin{equation} 
    m_\mathrm{cut} \approx 1.4 \times 10^{18}\,\mathrm{g} \, \left(\frac{t_\mathrm{age}}{10^{10}\,\mathrm{yr}}\right)^{-2/3}\,,
\end{equation}
and it does not depend on $\sigma$.

\begin{figure}[tb]
    \centering
    \includegraphics[width=\linewidth]{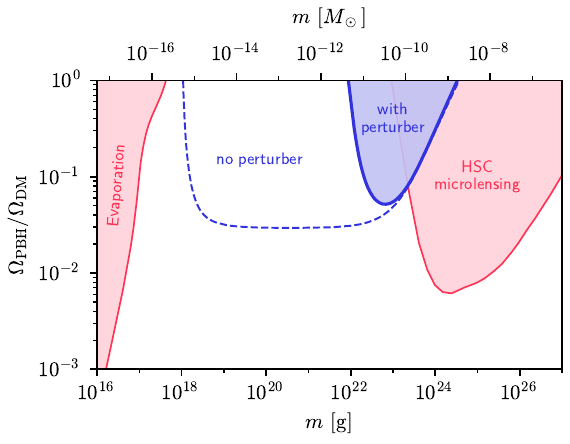}
    \caption{Impact of tidal perturbations on the proposed, hypothetical NS transmutation bounds on the PBH dark matter fraction. In keeping with previous work~\cite{CapelaConstraintsPBHDMNeutronStar2013}, we show the case for a DM density of $\rho=2\times 10^3\, \mathrm{GeV}/\mathrm{cm}^3$ in blue~\colorindicator{3333DD}, alongside present constraints on PBH dark matter~\cite{Iguaz:2021irx,Berteaud:2022tws,Smyth:2019whb} in red~\colorindicator{FF3355}. A monochromatic PBH mass function is assumed. Note that the blue region should not be interpreted as actual constraints, because the DM density inside globular clusters is not currently known with small enough uncertainties.} 
    \label{fig:constraints}
\end{figure}

In the past, it has been noted how constraints on PBH dark matter \textit{would} follow from the mere observation of old NS surviving in a dense DM environment, which was assumed to exist in globular clusters~\cite{CapelaConstraintsPBHDMNeutronStar2013}. To illustrate the impact of the perturber effect discussed in this article, we replicate the putative bounds from \cite{CapelaConstraintsPBHDMNeutronStar2013}, assuming a monochromatic PBH mass function, adopting the same fiducial parameters $\rho=2\times 10^3\, \mathrm{GeV}/\mathrm{cm}^3$ and velocity dispersion $\sigma = 4.0\,\mathrm{km/s} = 1.3 \times 10^{-5} c$. In Figure~\ref{fig:constraints} we show these hypothetical bounds (dashed blue curve), and how they would be affected accounting for stellar perturbations, with $n_p = 100 \,\mathrm{pc}^{-3}$ (solid blue curve). For comparison, the bounds from Hawking evaporation, based on isotropic~\cite{Iguaz:2021irx} and inner Galaxy~\cite{Berteaud:2022tws} hard x-ray observations, as well as microlensing with HSC~\cite{Smyth:2019whb}, are also included. While the curves in blue only represent potential sensitivities to PBH DM, since the DM density in such low velocity dispersion systems has not been firmly determined, this example shows the relevance of the tidal perturbation effects discussed here.

\section{Tidal Perturbations from Nearby Bodies} 

\subsection{Planetary Bodies} \label{sec:tidal_planets}

\begin{figure*}
    \centering
    \includegraphics[width=0.48\linewidth]{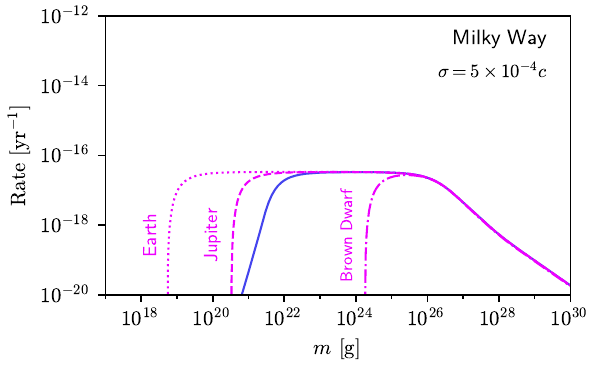}
    \includegraphics[width=0.48\linewidth]{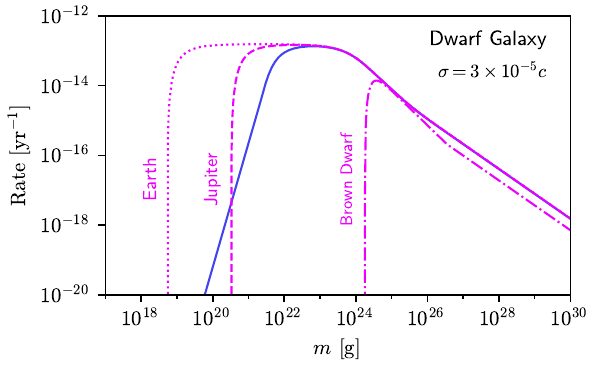}
    \caption{Impact on the transmutation rates of tidal perturbations due to NS companions of planetary masses: Earth-like ($M_p = M_\oplus$ and $R_p = 1\,\mathrm{au}$), Jupiter-like ($M_p = M_J$ and $R_p = 5.2\,\mathrm{au}$), and a brown dwarf/super-Jupiter in a very close orbit ($M_p = 0.01\,M_\odot$ and $R_p = 0.01\,\mathrm{au}$). 
    The curves can be interpreted as the expected suppression if every NS had a planet with those properties. If only a fraction $\mathcal{F}$ does, the suppression plateaus at $(1-\mathcal{F})$ times the maximum rate shown.}
    \label{fig:planet}
\end{figure*}

Despite being (surprisingly!) the first type of exoplanet discovered~\cite{1992Natur.355..145W}, planets orbiting NS are probably rare, based on current pulsar surveys~\cite{Nitu:2022ath}. While at present we do not believe that planetary perturbations should affect a significant fraction of the population of NS, possible selection effects in the surveys and the rapid evolution of exoplanetary science motivates us to provide here some first considerations of the potential disruption to merging PBH--NS systems brought by planetary bodies orbiting the NS.

Looking at the strength of the tidal effect, controlled by the quantity $Y$ defined in Eq.~\eqref{Ydef}, one may worry that catastrophic perturbations are induced by light bodies such as planets near the NS, given that
\begin{align}
    Y&=1.8\times 10^{-9} \,\frac{\mathrm{kg}}{\mathrm{m}^3}\frac{M_p}{M_\oplus}\frac{(1\,\mathrm{au})^3}{R_p^3}\nonumber\\
    &=6.8\times 10^{-20}\, \frac{\mathrm{kg}}{\mathrm{m}^3}\frac{M_p}{M_\odot}\frac{(1\,\mathrm{pc})^3}{R_p^3}\nonumber\,.
\end{align}
Indeed, the numerical benchmark value in the first line (obtained for an Earth-mass planet at 1 au orbital distance) is much larger than the benchmark in the second line (for a typical nearest stellar perturber). However, reasoning only in terms of the numerical values above is misleading, since the previous treatment has been assuming that the perturber lies far beyond the captured PBH orbit ($R_p\gg a$, see discussion around Eq.~\eqref{TidalForceApprox}), which is hardly the case for typical planetary orbits.

Fortunately, we can rely on the fact that the tidal perturbation due to a planet on a PBH initially captured by a NS on a wide orbit is extremely similar to the well-treated problem of planetary perturbations of cometary orbits around the Sun. Existing studies~\cite{LyttletonLossLongPeriodComets1964,YabushitaPlanetaryPerturbationOrbits1972,1999Icar..137...84W} prove that the orbital energy perturbation from a planet of mass $M_p$ is of the order
\begin{equation}
    \delta E \sim \frac{G M_p m}{R_p}\,,
\end{equation}
at least if the comet perihelion (or PBH periastron) is small with respect to the orbital radius of the planet, $R_p$, which is always the case when the PBH passes through the NS. As a rule of thumb, whenever $\delta E$ exceeds the binding energy $G M m / a$, we expect disruption of the PBH--NS system within a single orbit, which sets the condition
\begin{equation} \label{eq:planet_condition}
    a_1 < \frac{M}{M_p} R_p
\end{equation}
for initial captures potentially leading to merger in presence of a planet. This means that in the regime where contact interactions dominate the energy loss, the PBH mass must satisfy
\begin{equation} \label{eq:mPBH_planet_cutoff}
    m > 0.017 \frac{R}{R_p} M_p
\end{equation}
to allow most captured PBHs to eventually merge without immediate tidal disruption by the planet.

This effect, if unrealistically applied to the entire NS population, leads to the magenta curves in Figure~\ref{fig:planet}. It is shown for three example planetary bodies: Earth-like ($M_p = M_\oplus$ and $R_p = 1\,\mathrm{au}$), Jupiter-like ($M_p = M_J$ and $R_p = 5.2\,\mathrm{au}$), and a brown dwarf or super-Jupiter in a very close orbit\footnote{A strong perturber like the brown dwarf can also have some suppression effect at large masses, as seen in the GW capture dominated regime in the right panel of Figure~\ref{fig:planet}.} ($M_p = 0.01\,M_\odot$ and $R_p = 0.01\,\mathrm{au}$), similar to what has been observed in a number of pulsar systems \cite{Swihart:2022evu}. However, if the fraction of pulsars with planets is less than $\mathcal{O}(1\%)$, as current surveys seem to indicate, the effect when averaged over the whole population would of course be unnoticeable in the scale of Fig.~\ref{fig:planet}. In summary, we conclude that for the (likely small) fraction of NS which host planets, tidal effects may be comparable to the stellar tidal perturbations we have treated in the previous sections. 

Finally, for planetary perturbations of cometary orbits around the Sun, one can further rely on a statistical treatment of the cumulative kicks obtained over repeated orbits, yielding a timescale for orbital disruption~\cite{LyttletonLossLongPeriodComets1964,YabushitaPlanetaryPerturbationOrbits1972,1999Icar..137...84W}. However, this result cannot be translated to the PBH population relevant for this study, since the PBHs must come close to or cross the NS surface. The energy-loss effects they are subject to have no corresponding counterpart in the case of comets in the inner solar system, and the techniques used to study the latter are insufficient. We leave such an extension for future investigations. For the time being, we note that the results obtained in this section can then be considered conservative, since the actual cutoff mass could be actually higher. 

\subsection{Stellar Binary Companions}\label{sec:tidal_stars_bin}

\hl{Only a few percent of radio pulsars are found in binary systems~\cite{Tauris:2003pf}. Also, the relatively large number of known millisecond pulsars is due to a detection bias: millisecond pulsars are believed to constitute only a minor fraction of the total population of NS, dominated by old objects.  Under the reasonable assumption that the binary fraction of radio pulsars is representative of the whole population of NS, neglecting stellar companions of NS in our calculations should not lead to a significant error on the whole population.}
 
\hl{Yet, for the sake of completeness, it is worth noting that the} condition in Eq.~\eqref{eq:planet_condition} also applies to a stellar companion orbiting the NS at a distance $\ll 1\,$pc. \textit{For the fraction of NS in such close binary systems, the tidal perturbation may be so important that no competitive bounds can be expected from transmutation observables.} Based on Eq.~\eqref{eq:mPBH_planet_cutoff}, a binary companion of mass $M_p = 0.1\,M_\odot$ would have to orbit the NS at a radius $R_p \gtrsim 2.5~\mathrm{au}$ for constraints from NS transmutation to be competitive with those from microlensing. There is an exception, however: in the rare case where the companion of the NS is another NS on a very close orbit, the \hl{crossing rate} can be actually \textit{enhanced} with respect to the ones computed here. The enhancement factor has been estimated to reach about 4 for binaries with orbital periods of 4 to 8 hours~\cite{BrayeurEnhancementDMCapture2012}. While conceptually interesting, this effect is negligible for a population of NS, whose fraction in such close binaries is very small.

\section{Discussion} \label{sec:discussion}

Following the discovery of gravitational waves from coalescing binary black holes, the possibility that primordial black holes (PBHs) formed in the early universe has been subject to renewed and detailed scrutiny. Unlike astrophysical black holes, PBHs could form with masses much smaller than a solar mass, with lighter objects forming at earlier times in the history of the universe. A particularly exciting possibility is that sub-lunar PBHs may have formed much earlier than the electroweak symmetry breaking era, still an uncharted territory as far as the predominant physics is concerned and on which we have no cosmological information. It is then more easily conceivable that the rather extreme conditions needed to form PBHs may be met without violating existing constraints. In the so-called asteroid-mass window, provided that the PBH mass function is not too wide~\cite{Gorton:2024cdm}, such objects may constitute the totality of the dark matter in the universe, providing a qualitatively different alternative to particle dark matter scenarios.

These PBHs turn out to be difficult to probe. Yet, a number of proposals have been put forward, pointing to signatures related to the capture of these objects into compact stars, notably neutron stars. Spectacular consequences are expected in particular if the PBHs settle into neutron stars and cause their transmutation into black holes within timescales shorter than a few Gyr. We have studied how the capture process is affected by the tidal perturbation of nearby stellar or planetary objects. We find that such a disturbance prevents the lightest viable PBH dark matter candidates from efficiently settling into the star and causing a transmutation. Effectively, that makes PBHs of masses between $\sim 10^{18} - 10^{22}\,\mathrm{g}$ harder to probe than naively thought. Our main result is summarized in Eqs.~\eqref{eq:m_pert} and \eqref{eq:Gamma_pert}, describing respectively the critical mass below which tidal perturbations are relevant and the suppressed rate below it. The methodology and formulae of this paper can be readily extended to other celestial bodies.

A somewhat related question is the impact that other PBHs have on the capture. For a typical Galactic DM density of $0.5\,\mathrm{GeV/cm}^3\simeq 0.01 \, M_\odot/\mathrm{pc}^3$, the total enclosed DM mass within the distance of the nearest star is comparable to or smaller than the mass of the nearest stellar perturber, so it should not contribute more than the stars. However, for NS without planets that are surrounded by high DM densities, it is possible that nearby PBHs may provide the leading perturber's role. Depending on the parameters, the treatment adopted for stellar or planetary perturbers (or even a stochastic treatment) can be appropriate to treat these cases. Some differences could however be expected, particularly since the perturber would have a mass comparable to the captured PBH and much smaller than the target NS. Another question of interest is to assess how important are genuine three body interactions in capture and disruption processes for PBHs. We leave the investigation of such topics to future research.

\begin{acknowledgments}
This work was supported by the UChicago--CNRS Collaboration Program ``New Ideas in the Search for Dark Matter''. I.H. was also supported by the Brinson Foundation.
\end{acknowledgments}

\appendix

\section{Motion of PBH inside NS} \label{appendix:interior_orbits}

Once the PBH enters the interior of the NS, the gravitational potential changes, leading to deviations from a Keplerian orbit. If we assume that the NS has a constant density profile, then this results in a harmonic potential with $F_g \propto r$. In this case, the orbits are still elliptical but they are centered at $r=0$. The true behavior for a realistic NS density profile will lie somewhere in between this central elliptical orbit and a standard Keplerian orbit.

The semi-major axis $\alpha$ and semi-minor axis $\beta$ of the interior orbit can be written in terms of the axes $a$ and $b$ of the exterior orbit:
\begin{align}
    \alpha &= R \sqrt{\left( \frac{3}{2} \pm \frac{R}{2a} \right) + \sqrt{\left( \frac{3}{2} \pm \frac{R}{2a} \right)^2 - \frac{b^2}{a R}}}\\
    \beta &= R \sqrt{\left( \frac{3}{2} \pm \frac{R}{2a} \right) - \sqrt{\left( \frac{3}{2} \pm \frac{R}{2a} \right)^2 - \frac{b^2}{a R}}},
\end{align}
where the $+$ sign applies to hyperbolic exterior orbits, and the $-$ sign to elliptical ones. The eccentricity can be written $\varepsilon = \sqrt{1 - (\beta /\alpha)^2}$. The elliptical trajectory through the interior of the NS in polar coordinates is then
\begin{equation}
    r_\mathrm{in}(\phi) = \frac{\alpha \beta}{\sqrt{\beta^2 + (\alpha^2 - \beta^2) \cos^2 \phi}}\, .
\end{equation}

\begin{figure}[tb]
    \centering
    \includegraphics[width=\linewidth]{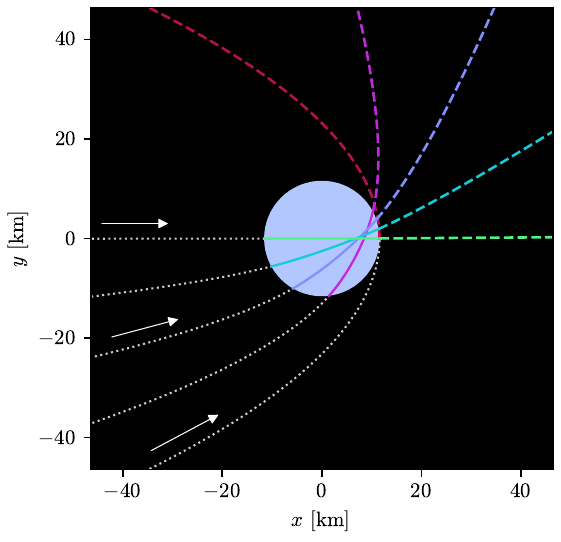}
    \caption{Trajectory of the first pass of the PBH through the NS for several values of $b_0$. The white dotted lines show the initial hyperbolic trajectories, the solid colored lines show the non-Keplerian interior trajectories, and the dashed colored lines show the resulting precessed orbits. The lines have $b_0$ values ranging from $0$ to $b_c$ in steps of $0.25 b_c$.}
    \label{fig:interior_orbits}
\end{figure}

There are two important consequences of this altered trajectory. First, the PBH speed at closest approach is lower, leading to slightly decreased energy losses. Second, it induces an orbital precession during each successive pass, as shown in Figure~\ref{fig:interior_orbits}. As noted in the main text, 
the changing orientation of the orbit complicates the consideration of tidal effects over many orbits. To quantify the amount of precession, the shift in the argument of periapsis $\omega$ during each pass through the star is
\begin{equation}
\begin{split}
    \Delta \omega = 2 \Bigg[ & \arccos \left( \frac{\beta}{R} \sqrt{\frac{\alpha^2 - R^2}{\alpha^2 - \beta^2}} \right) \\ & -  \arccos \left( \frac{b^2 - a R}{R \sqrt{a^2 + b^2}} \right) \Bigg]\,.
\end{split}
\end{equation}
See \cite{GenoliniRevisitingPBHCapture2020} for additional discussion of this effect.

\section{Energy Loss Mechanisms} \label{appendix:e_loss}

In this Appendix we review the interaction mechanisms and the resulting drag force a PBH undergoes while crossing a NS. The associated energy loss $\Delta E$ and angular momentum loss $\Delta L$ used throughout this paper are given by the following integrals:
\begin{align}
    \Delta E &= \int_\mathcal{C} (\vec{v}\cdot\vec{F}) \, \dd t\,, \label{eq:Delta_E_def}\\
    \Delta \vec{L} &= \int_\mathcal{C} (\vec{r} \times \vec{F}) \,\dd t \, ,
\end{align}
where $\cal C$ is the PBH trajectory near/inside the NS (far away, the relative contributions to $\Delta E$ and $\Delta \vec{L}$ are negligible). As mentioned in the paper, these calculations are performed within the non-relativistic approximation, which should work well, since general relativistic corrections on the speed can reach at most 5\%, see Ref.~\cite{GenoliniRevisitingPBHCapture2020}. Furthermore, we adopt benchmark values for the NS mass ($M = 1.52 \,M_\odot$) and radius ($R = 11.6 \, \mathrm{km}$) based on the BSK-20-1 equation of state~\cite{PotekhinAnalyticalRepresentationsUnified2013}. For simplicity, we assume a constant density profile inferred from these parameters, which is used to compute the drag experienced by the PBH as it traverses the NS. The different drag forces are reviewed in~\cite{GenoliniRevisitingPBHCapture2020} and briefly recalled here.\\

\textbf{Dynamical friction} --- When the PBH crosses the NS its motion is mostly supersonic \cite{GenoliniRevisitingPBHCapture2020,Baumgarte:2024iby}. Hence it is dragged by the gravitational pull from the neutron wake---so-called dynamical friction---in the collisionless regime, which is written~\cite{ChandrasekharBrownianMotionDynamical1949,BinneyGalacticDynamics1987,OstrikerDynamicalFrictionGaseous1999}:
\begin{equation}
\vec{F}_\mathrm{dyn}=-4\pi G^2 m^2\rho \ln{\Lambda_\mathrm{dyn} }(v) \frac{\boldsymbol{v} }{v^3}\;,\label{eq:df}
\end{equation}
where $\rho$ is the NS medium density. Note that in this expression the \textit{Coulomb logarithm} is computed following \cite{CapelaConstraintsPBHDMNeutronStar2013}, taking into account the degenerate nature of the neutron fluid.

\textbf{Accretion} --- Neutrons sufficiently close to the PBH trajectory, i.e. for impact parameters below $d_\mathrm{crit}$, are accreted at rest. Thus the PBH is also dragged by this process. It was shown \cite{CapelaConstraintsPBHDMNeutronStar2013} that the corresponding drag force takes the form:
\begin{equation}
\vec{F}_\mathrm{acc}=-\pi d_\mathrm{crit}^2 \rho \gamma^2  v\boldsymbol{v} \;,\label{eq:acc}
\end{equation}
where $\gamma$ is the Lorentz factor, and $d_\mathrm{crit}$ is the \textit{critical} impact parameter below which neutrons fall into the BH.

\textbf{Surface waves} --- Passing through the NS, the PBH excites transverse density waves or surface waves. Using a simple incompressible fluid model, the energy loss has been estimated in Ref.~\cite{DefillonTidalCapturePBHs2014} to be of the order:
\begin{equation}
|\Delta E_\mathrm{surf}| \sim \frac{3\,G\,m^2}{R}\;.\label{eq:swel}
\end{equation}

\textbf{Gravitational waves} ---
By passing close to the NS, the PBH motion is also damped by gravitational wave (GW) emission. The power emitted takes the general form \cite{PetersGravitationalRadiationMotion1964}:
\begin{equation}
\frac{\dd E}{\dd t} = \frac{G}{5\,c^5} \dddot{Q}_{ij} \dddot{Q}_{ij} \,,\label{PowGW}
\end{equation}
$Q_{ij}$ being the traceless quadrupole moment tensor. Note that this is basically the only energy loss mechanism that does not require contact between the PBH and the NS. For the case of a hyperbolic encounter where there is no contact, the energy loss takes the form:
\begin{equation} \label{eq:DeltaEGW}
    |\Delta E_\mathrm{gw}| \approx \frac{170 \pi G^7 M^6 m^2}{3 c^5 \, b_0^7 \, v_0^7} \, ,
\end{equation}
where we have assumed that $a_0 \gg b_0$ (the parabolic $e \approx 1$ limit).
The generalization of this expression to the case where the PBH enters the NS can be found in Appendix 2 of \cite{GenoliniRevisitingPBHCapture2020}, considering a constant NS density profile for the NS. The derivation of a more general radiation-reaction force from GW emission can be found in Appendix B.1 of \cite{Baumgarte:2024iby} which takes into account motion within a NS with varying density profile.

\begin{figure}[tb]
    \centering
    \includegraphics[width=\linewidth]{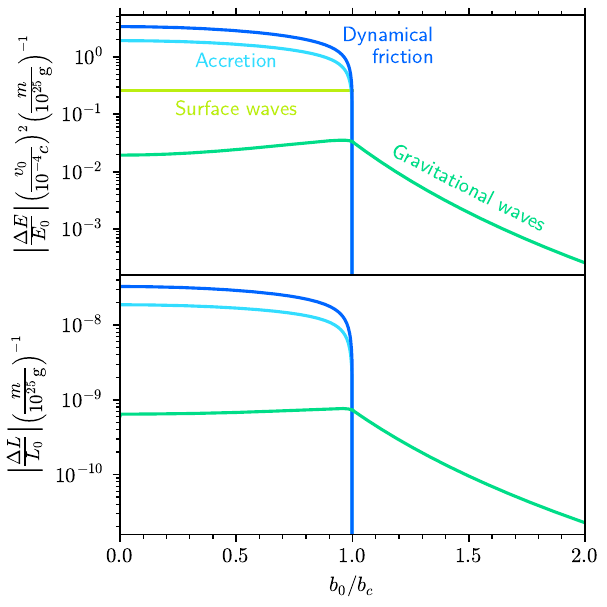}
    \caption{Relative energy and angular momentum losses as a function of initial impact parameter $b_0/b_c$, for each of the four dissipative mechanisms. They have been scaled to remove the leading dependence on $m$ and $v_0$.}
    \label{fig:energy_losses}
\end{figure}

Summary plots of the relative energy and angular momentum losses are shown in Figure~\ref{fig:energy_losses}. The dominant source of energy loss as the PBH crosses the NS is dynamical friction, closely followed by accretion, which is smaller by a factor of approximately 2. Surface waves and gravitational wave emission contribute sub-dominantly. Outside the NS, gravitational waves are the sole mechanism for energy loss, with a steep dependence on the initial velocity, scaling as $\Delta E_\mathrm{gw} \propto v_0^{-7}$. That dependence highlights the critical role of gravitational wave emission in systems with low velocity dispersion. Furthermore, note that the change in angular momentum from these dissipative interactions both inside and outside the NS are always substantially smaller than $|\vec{L}_0|$, which justifies the approximation $\vec{L}_1\approx \vec{L}_0$.

Apart from initiating the capture of the PBH, these processes also govern its subsequent post-capture dynamics. This phase of motion has recently been revisited using a general relativistic (GR) framework in Ref.~\cite{Baumgarte:2024iby}, where the BH is treated as a point mass moving within a realistic, radially dependent density profile derived from a polytropic equation of state. This more sophisticated treatment yields the following differences compared to earlier, simpler models:
(i) During the inspiral phase, these forces are largely ineffective at circularizing the PBH orbit, particularly for low-mass PBHs (see Fig.~10 in Ref.~\cite{Baumgarte:2024iby});
(ii) As a consequence, the GW signal is altered: its amplitude becomes modulated due to orbital precession, leading to quasiperiodic beating patterns (see also Ref.~\cite{Baumgarte:2024mei});
(iii) The scaling relation for the orbital semi-major axis as a function of PBH mass is modified: Eq.~(62) of \cite{Baumgarte:2024iby} gives a power-law index of 0.83, contrasting with the value of 0.50 obtained in Eq.~(37) of \cite{GenoliniRevisitingPBHCapture2020}.
On the other hand, the altered Bondi accretion for stiff EoS \cite{Richards:2021zbr,Richards:2021upu} should not lead to  marked differences with respect to the approximation used here.

\section{Analytic Rate Approximations} \label{appendix:analytic}

\begin{figure}[t]
    \centering
    \includegraphics[width=\linewidth]{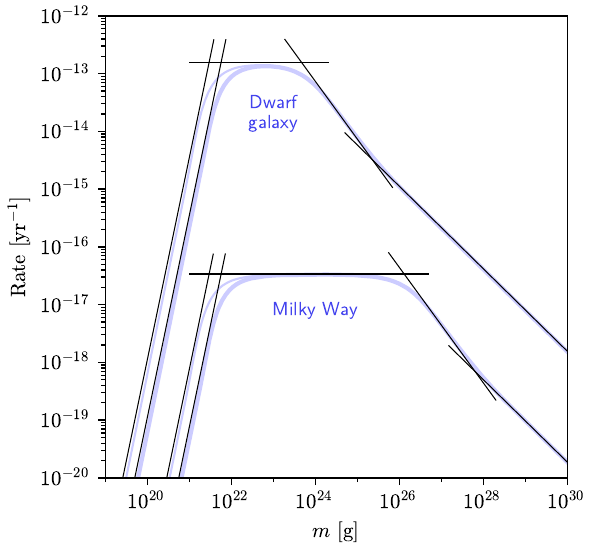}
    \caption{Same as the ``merger (with perturber)'' lines from Figure~\ref{fig:rates}, but with analytic approximations to the four regimes from Eqs. \eqref{eq:Gamma_gw_full}, \eqref{eq:Gamma_cross_full}, \eqref{eq:Gamma_cap_full}, and \eqref{eq:Gamma_pert_full} overlaid in black.}
    \label{fig:analytics}
\end{figure}

Here we show the derivation of the merger rate in each of the regimes mentioned in Section~\ref{sec:effects}, describing the assumptions made. Figure~\ref{fig:analytics} compares these analytic expressions to the full numerical results, showing that they are indeed accurate.

The gravitational wave capture rate is calculated by considering the energy loss $\Delta E_\mathrm{gw}$ defined in Eq.~\eqref{eq:DeltaEGW}. We can define
\begin{equation}
    b_\mathrm{gw} \equiv \left(\frac{340\pi \, G^7 M^6 m}{3 c^5 v_0^9}\right)^{1/7}
\end{equation}
as the largest value of $b_0$ for which capture is possible (where $|\Delta E_\mathrm{gw}| > m v_0^2 / 2$). Performing the rate integrals over $b_0$ and $v_0$ gives
\begin{equation} \label{eq:Gamma_gw_full}
\begin{split}
    \Gamma_\mathrm{gw} &= \frac{\rho}{m} \int_0^\infty \dd v_0 \, p(v_0) \,  v_0 \left( \pi b_\mathrm{gw}^2 \right)
    \\
    &\approx 
    \sqrt{2\pi} \, \mathsf{\Gamma}(\tfrac{5}{7}) \, \frac{\rho \, \sigma}{m} \left(\frac{170 \pi G^7 M^6 m}{3 c^5 \sigma^9}\right)^{2/7},
\end{split}
\end{equation}
where $\mathsf{\Gamma}(n)$ is the standard gamma function.

The crossing rate is calculated in a similar way, except we restrict to initial trajectories that pass through the NS, so the upper limit on the $b_0$ integral is simply $b_c$. Additionally, we assume that $v_0^2 \ll \sqrt{G M / R}$, which is effectively an assumption that the PBH velocity distribution is non-relativistic. Given the definition of $b_c$ in Eq.~\eqref{eq:bc}, we can approximate
\begin{equation}
    b_c \approx \frac{\sqrt{2 G M R}}{v_0} \, .
\end{equation}
Thus we find the rate
\begin{equation} \label{eq:Gamma_cross_full}
\begin{split}
    \Gamma_\mathrm{cross} &= \frac{\rho}{m} \int_0^\infty \dd v_0 \, p(v_0) \,  v_0 \left( \pi b_c^2 \right)
    \\
    &\approx \frac{\sqrt{8\pi} \, G M R \, \rho}{\sigma m} \, .
\end{split}
\end{equation}

To calculate the capture rate from direct contact interactions, we first find an approximate analytic expression for these energy losses. For drag forces of the form $\vec{F} = -A \vec{v}$, we can integrate Eq.~\eqref{eq:Delta_E_def} analytically. Eqs.~\eqref{eq:df} and \eqref{eq:acc} can be approximately expressed in this form, with 
\begin{equation}
    A = \frac{4\pi G^2 m^2 \rho_\mathrm{NS} \ln\Lambda_\mathrm{eff}}{v^3} \approx \ln\Lambda_\mathrm{eff}\sqrt{\frac{G}{3 M R^3}} \,m^2 \, ,
\end{equation}
where for $v$ we have used the interior trajectory's periapsis velocity $v_\mathrm{max} \approx \sqrt{3 G M / R}$, and we defined an effective Coulomb logarithm for the dominant sources of energy losses, dynamical friction and accretion:
\begin{equation}
    \ln\Lambda_\mathrm{eff} \equiv \ln\Lambda_\mathrm{dyn} + \frac{\gamma^2 v^4 d_\mathrm{crit}^2}{4 G^2 m^2} \approx 16.
\end{equation}
Although $\ln\Lambda_\mathrm{eff}$ also has a mild dependence on PBH velocity, we evaluate it at the constant velocity $v_\mathrm{max}$ as well, giving $\ln\Lambda_\mathrm{eff} \approx 16$. In calculating $\Delta E$, we again take the $v_0^2 \ll \sqrt{G M / R}$ limit, and because the dependence on $b_0$ is not especially strong, we also assume that $b_0 \ll b_c$. We finally arrive at
\begin{equation} \label{eq:deltaE_approx}
    |\Delta E| \approx \Big(\sqrt{\tfrac{2}{3}} + \sqrt{3} \arctan\tfrac{1}{\sqrt{2}}\Big) \ln\Lambda_\mathrm{eff} \frac{G m^2}{R}.
\end{equation}
Note that this procedure has removed all dependence on $b_0$ and $v_0$. The expression obtained here is consistent with the full calculation shown in Figure~\ref{fig:energy_losses}.

Requiring $|\Delta E| > m v_0^2 / 2$, the condition for PBH capture is then $v_0 < v_\mathrm{cap} \equiv \sqrt{2|\Delta E| \, / \, m}$. Since we are in the regime where direct contact interactions dominate, we still require the crossing condition of $b_0 < b_c$, but now there is also an upper limit on the $v_0$ integral:
\begin{equation} \label{eq:Gamma_cap_full}
\begin{split}
    \Gamma_\mathrm{cap} &= \frac{\rho}{m} \int_0^{v_\mathrm{cap}} \dd v_0 \, p(v_0) \,  v_0 \left( \pi b_c^2 \right)
    \\
    &= \frac{\sqrt{8\pi} \, G M R \, \rho}{\sigma m} \left( 1 - e^{-v_\mathrm{cap}^2 / (2\sigma^2)} \right)
    \\
    \Gamma_\mathrm{plat} &\approx \sqrt{8 \pi} \Big(\sqrt{\tfrac{2}{3}} + \sqrt{3} \arctan\tfrac{1}{\sqrt{2}}\Big)\frac{\ln\Lambda_\mathrm{eff} \, G^2  M \rho}{\sigma ^3}\, .
\end{split}
\end{equation}
In the last line, we have Taylor expanded the exponential. This is because we are interested in the regime where most encounters do not lead to capture, which occurs when $v_\mathrm{cap} \ll \sigma$. Figure~\ref{fig:analytics} shows that this is an excellent approximation for the purpose of calculating the capture rate. Note that if we instead took the limit of $v_\mathrm{cap} \gg \sigma$ in Eq.~\eqref{eq:Gamma_cap_full}, we would precisely get the crossing rate in Eq.~\eqref{eq:Gamma_cross_full}, simply because in that regime, practically every encounter leads to capture.

We now discuss the merger rate in the presence of tidal perturbers, the definition of which was initially given in Eq.~\eqref{eq:rate}. We first note that the magnitude of the angular momentum perturbation from Eq.~\eqref{eq:delta_L} can be written in the $e_1 \approx 1$ limit as
\begin{equation}
\begin{split}
    |\delta \vec{L}| = {} & 15 \pi m \sqrt{\frac{G}{M}} a_1^{7/2} Y \\ & \times \sqrt{\cos^2\chi \sin^2\psi (\cos^2\psi + \sin^2\chi \sin^2\psi)} \, .
\end{split}
\end{equation}
We will approximate 
\begin{equation} \label{eq:a1_approx}
    a_1 \approx \frac{G M m}{2 |\Delta E|}
\end{equation}
because often $|\Delta E| \gg E_0$. Note that in the limits taken, $|\delta \vec{L}|$ does not depend on $b_0$ or $v_0$.

It turns out that the criterion for successful merger is most frequently failed when the angular momentum perturbation becomes larger than the initial angular momentum, or $|\delta \vec{L}| > m v_0 b_0$. In other words, the captured orbit is disrupted when $b_0 < b_\mathrm{dis} \equiv |\delta \vec{L}| / (m v_0)$. Rather than setting this as a bound on the $b_0$ integral, we instead impose this condition on the integrals over the angular variables $\chi$ and $\psi$. Then, the integrals over $b_0$ and $v_0$ can be set up identically to $\Gamma_\mathrm{cap}$, yielding the same result as Eq.~\eqref{eq:Gamma_cap_full}.

To evaluate the remaining integrals, we observe that for fixed values of $v_0$ and $Y$, if $b_\mathrm{dis} > b_c$, then there exist no values of $\chi$ or $\psi$ where the PBH is simultaneously \textit{captured} and also \textit{not disrupted}. Thus, we require 
\begin{equation}
    b_\mathrm{dis} = \frac{|\delta \vec{L}|}{m v_0} < \frac{\sqrt{2\, G M R}}{v_0} = b_c\,.
\end{equation}
Conveniently, the $v_0$ dependence cancels here. Thus we can recast this inequality into a condition on $\chi$ and $\psi$:
\begin{equation} \label{eq:chipsi_condition}
\cos^2\!\chi \sin^2\!\psi \, (\cos^2\!\psi + \sin^2\!\chi \sin^2\!\psi) < \frac{2 M^2 R}{(15\pi)^2 Y^2 a_1^7}
\end{equation}
We do not elaborate on the process of performing the $\chi$ and $\psi$ integrals here, but the result is an additional factor in the rate of
\begin{equation*}
\frac{1}{4\pi} \! \int\limits_0^{2\pi} \!\dd\chi \int\limits_0^\pi \!\dd\psi \sin\psi ~ \Theta(b_c - b_\mathrm{dis}) \approx 
\left( \frac{4}{\pi} - 1 \right) \frac{\sqrt{2 R} M}{15 \pi Y a_1^{7/2}}\,.
\end{equation*}
Intuitively, the rate of merger decreases as the strength of the tidal perturber (controlled by $Y$) increases.
Only the integral over $Y$ remains, the result of which is
\begin{equation}
    \int\limits_0^\infty \dd Y \, p(Y) \, Y^{-1} = \frac{3 (\gamma -1)}{4 \pi  \gamma  M_\mathrm{min} n_p}\,,
\end{equation}
where we assumed $M_\mathrm{min} \ll M_\mathrm{max}$.

Combining the above expressions, we find
\begin{align} \label{eq:Gamma_pert_full}
    \Gamma_\mathrm{pert} \approx {} & \frac{8}{5\pi} \sqrt{\frac{2}{\pi }} \left(\frac{4}{\pi} - 1\right) \Big(\sqrt{\tfrac{2}{3}} + \sqrt{3} \arctan\tfrac{1}{\sqrt{2}}\Big)^{9/2} \nonumber
    \\ &\times \ln\Lambda_\mathrm{eff}^{9/2} \, \frac{\gamma-1}{\gamma} \,  \frac{G^2 m^{7/2} \rho}{M^{3/2} R^3 \sigma^3 M_\mathrm{min} n_p}\,.
\end{align}
The PBH mass for which the angular $\chi-\psi$ space begins to become restricted by the condition in Eq.~\eqref{eq:chipsi_condition} is
\begin{equation}
    m_\mathrm{pert} =  \frac{\left( \left(\frac{5\pi^2}{4}\right)^2 \left(\frac{\gamma}{\gamma-1}\right)^2 M^5 R^6 M_\mathrm{min}^2 n_p^2 \right)^{1/7}}{\ln\Lambda_\mathrm{eff} \Big(\sqrt{\frac{2}{3}} + \sqrt{3} \arctan\tfrac{1}{\sqrt{2}}\Big)}\,.
\end{equation}

\bibliographystyle{utphys}
\bibliography{references}

\end{document}